\documentclass[%
 reprint,
 amsmath,amssymb,
 aps,
 prb
]{revtex4-2}

\usepackage[utf8]{inputenc}
\usepackage[T1]{fontenc}
\usepackage[version=4]{mhchem}
\usepackage{graphicx}
\usepackage[caption=false]{subfig}
\usepackage{standalone}
\usepackage{amsmath}
\usepackage{amsfonts}
\usepackage{amssymb}

\usepackage[linktocpage=true,
  colorlinks=true, 
  pdfborder={0 0 0}, 
  linkcolor=blue,
  citecolor=red,
  filecolor=yellow,
  urlcolor=blue,
  bookmarks,
  pdfauthor={},
]{hyperref}

\begin{document}
\title{Origin of Anomalous Hall effect in Cr-doped \ce{RuO2}}

\author{Andriy Smolyanyuk}
\email{andriy.smolyanyuk@tuwien.ac.at}
\affiliation{Institute of Solid State Physics, TU Wien, 1040 Vienna, Austria}
\author{Libor \v Smejkal}
\affiliation{Max Planck Institute for the Physics of Complex Systems, N\"othnitzer Str. 38, Dresden, Germany}
\affiliation{Max Planck Institute for Chemical Physics of Solids, N\"othnitzer Str. 40, Dresden, Germany}
\affiliation{Institute of Physics, Czech Academy of Sciences, Cukrovarnick\'a 10, 162 00, Praha 6, Czech Republic}
\author{Igor I. Mazin}
\email{imazin2@gmu.edu}
\affiliation{George Mason University, 22030 Fairfax, USA}

\date{\today}

\begin{abstract}
\ce{RuO2} is one of the most highlighted candidates for altermagnetism.
However, the most recent muon spin spectroscopy and neutron studies demonstrated the absence of magnetic order in this system.
The electronic structure of \ce{RuO2} hints at a possibility of realizing a magnetically ordered state upon hole doping, and such a possibility was explored experimentally in Cr-doped \ce{RuO2}, where it was suggested that this system exhibits the anomalous Hall effect (AHE) due to altermagnetism.
In this manuscript, based on our density functional calculations, we revise the results obtained for this system and propose a different interpretation of experimental results.
Our calculations suggest that extra holes are bound to Cr impurity and do not dope Ru bands, which remain nonmagnetic.
Thus, the observed AHE is not due to the altermagnetism but stems entirely from magnetic Cr ions.
\end{abstract}

\maketitle

\section{Introduction}
Altermagnetism (AM) is a recently proposed third type of collinear magnetism~\cite{smejkal_beyond_2022} and is currently gaining much attention in the scientific community.
In AM systems, due to their symmetry, there is no net magnetization, but time-reversal symmetry is broken in both direct and reciprocal space, and depending on the spin axes, symmetry allows for anomalous Hall effect (AHE)~\cite{smejkal_crystal_2020}.
Part of the attention is due to the potential applications in spintronics: metallic AM would imply the possibility of giant/tunneling magnetoresistance effects or the spin-polarized current generation in this systems~\cite{smejkal_giant_2022, smejkal_emerging_2022,gonzalez-hernandez_efficient_2021,shao_spin-neutral_2021}, which could be utilized to construct magnetic memory devices.

One of the most discussed candidates in the context of altermagnetism is \ce{RuO2}.
While originally thought to be a Pauli paramagnet~\cite{guthrie_magnetic_1931, ryden_magnetic_1970,fletcher_magnetic_1968}, five years ago neutron scattering studies suggested that the actual ground state was antiferromagnetic, albeit with a very small magnetic moment of $\sim 0.05\ \mu_B$/Ru~\cite{berlijn_itinerant_2017}.
Despite the ordered moment being so small, the material has sparked a lot of interest recently because the claimed magnetic structure was altermagnetic~\cite{smejkal_crystal_2020}, and the material is a good metal. Subsequently, several dozen experimental and theoretical papers claiming altermagnetism in \ce{RuO2} have appeared.

There have been several red flags related to this compound that should have been studied but hitherto were not.
Indeed, the reported from polarized neutron scattering magnetic moment was not consistent with unpolarized spectra.
Moreover, unadulterated density functional (DFT) calculations did not give any magnetic ground state, not even close, and it is unheard of that DFT, a mean-field theory, would underestimate the tendency to magnetism in itinerant systems~\cite{berlijn_itinerant_2017,smolyanyuk_fragility_2024}.
Indeed, it was finally experimentally demonstrated that no magnetic order is present in this system~\cite{hiraishi_nonmagnetic_2024, kessler_absence_2024}, while the old neutron data~\cite{berlijn_itinerant_2017} were misinterpreted and are actually consistent with the paramagnetic state.

This finding shifted the interest into the possibility of artificially driving \ce{RuO2} towards altermagnetism by doping, strain, or morphologically.
In particular, it was predicted that \ce{RuO2} should undergo a transition into an (alter)magnetic state upon a hole doping~\cite{smolyanyuk_fragility_2024}.
Recently, an experimental study reported that \ce{RuO2} with 20\% Cr doping is magnetic (presumed altermagnetic)~\cite{wang_emergent_2023}. This study was not informed of the latest findings~\cite{hiraishi_nonmagnetic_2024,kessler_absence_2024} and based their interpretation upon the (now proved to be incorrect) assumption that \ce{RuO2} is antiferromagnetic with the easy axis [001] and interpreted their results as rotating of the N\'eel vector due to Cr doping, not as inducing magnetism in otherwise nonmagnetic compound. Obviously, in view of the new findings, this experimental result needs to be revisited.

In this manuscript, based on the results of density functional calculations, we can confirm that doping of 0.4 hole (Cr has two $d$-electrons fewer than Ru) is sufficient to drive \ce{RuO2} magnetic.
However, we also find that the extra holes remain bound to Cr impurity and do not dope the Ru bands, so Ru remains locally nonmagnetic (apart from the normal metallic screening effect).
Furthermore, statistically with probability of 83\% a random Cr will have at least one other Cr as a nearest neighbor, and there is a 50\% probability that it belongs to a cluster of three or more Cr.
Given that rutile \ce{CrO2} is a known strong ferromagnet (FM), there is a possibility, even likeliness, that the AHE observed in Ref.~\onlinecite{wang_emergent_2023} has nothing to do with still elusive altermagnetism of \ce{RuO2}, but rather directly associated with Cr magnetic ions.

\section{Results and discussion}
\subsection{DFT results}
\begin{figure*}[t!]
     \includegraphics[width=\textwidth]{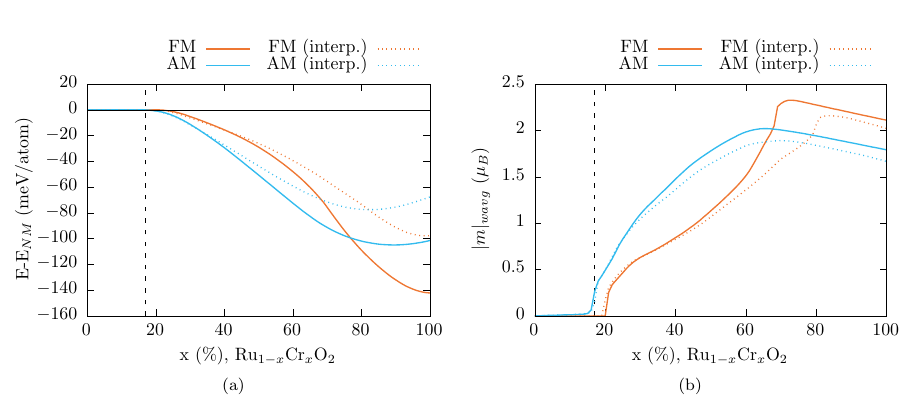}
     \caption{Energies (a) and weighted average of local magnetic moments (b) of various magnetic phases of Cr-doped \ce{RuO2} in the VCA approximation.
     Energies are given w.r.t the energy obtained from non-spin-polarized calculation.
     Solid lines correspond to calculations using the experimental crystal structure of \ce{RuO2} and dashed lines using the crystal structure obtained from the linear interpolation between \ce{RuO2} and \ce{CrO2} crystal structures depending on the Cr concentration.
     }
     \label{fig:VCA}
\end{figure*}

Here, the aim is to check whether the Cr-doped \ce{RuO2} is an altermagnet.
To probe the effect of such doping in the itinerant regime and to check if the correct magnetic ground state is obtained we employed, first, the virtual crystal approximation (VCA).
The result is plotted in Fig.~\ref{fig:VCA}, solid lines: the system becomes magnetic at the concentration of Cr at about 17\% with the altermagnetic state being more favorable than the ferromagnetic one.
At the concentrations of Cr about 80\%, the transition to the ferromagnetic state happens, which is in agreement with experimental data for pristine \ce{CrO2}, which is ferromagnetic.

\begin{figure*}
    \includegraphics[width=\textwidth]{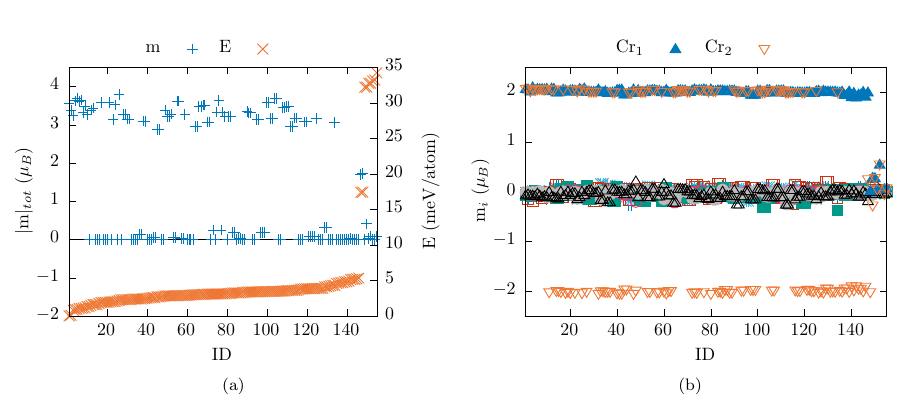}    
    \caption{The absolute value of total magnetic moment with corresponding energy (a) and local magnetic moment (b) for a set of supercells of \ce{RuO2} with 20\% Cr doping;
    each cell contains 2 Cr atoms.
    In panel (a) blue '+' symbols and orange 'x' symbols denote the total magnetic moment in the cell and the energy per atom w.r.t the lowest state.
    In panel (b) filled blue triangles and open orange triangles denote the local magnetic moment on the respective Cr atom, while
    other symbols denote the same quantity for Ru atoms.
    When the ferromagnetic state is realized, the magnetization direction is assumed to be positive.
    }
    \label{fig:SC}
\end{figure*}

To make this check more realistic and to see the effect of crystal structure change, the same calculations were performed, but with lattice constants and oxygen position linearly interpolated between those of pristine \ce{RuO2} and \ce{CrO2}, leading to the same qualitative result, but the AM-FM transition shifted to the slightly higher Cr concentration (see the dashed lines in Fig.~\ref{fig:VCA}).

To further test if the previous approximations are realistic, we performed a series of supercell calculations for the \ce{RuO2} with 20\% Cr doping.
In total, we used 52 symmetry inequivalent supercells (including non-diagonal variants) that have a volume five times that of the pristine \ce{RuO2} and each containing two Cr atoms.
The experimental lattice parameters for the parent pristine cell were used~\cite{cotton_crystal_1966, sd_0541479}.
For each supercell, three different magnetic configurations were investigated: ferromagnetic, altermagnetic (note that in this case, both Cr atoms may have the same orientation of magnetic moment due to their positions), and where only Cr atoms were assigned with the initial magnetic moment, but in opposite directions.

The results of these calculations are highlighted in Fig.~\ref{fig:SC}a, where the calculations are sorted by the energy.
The first important finding is that the 10 lowest energy structures are actually ferromagnetic.
The second is that the magnetic moments on Ru atoms are not uniformly distributed (as would be in the case of the itinerant ansatz) and that the main contribution to the total magnetization is due to Cr atoms (see Fig.~\ref{fig:SC}b).
The latter suggests that the magnetic moment on Ru atoms is induced by the localized moments of Cr atoms, rather than being an effect of hole doping.

\begin{figure*}
    \centering
    \includegraphics[width=\textwidth]{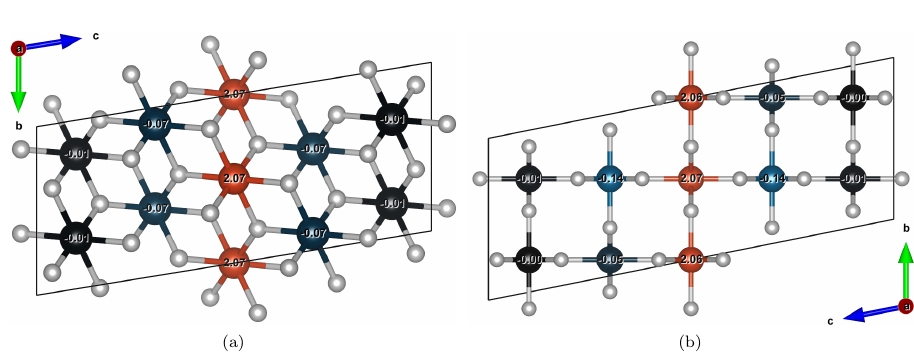}    
    \caption{Magnetic moment distribution for the two lowest energy structures.
    The color is linearly interpolated based on the value of magnetic moment from black to red or blue for spin up or down, where the highest color value is assigned to 0.2 $\mu_B$ (thus, the magnetization of Cr will always be painted with the interpolation edge color).
    The values printed on each atom correspond to the local magnetic moment.
    For convenience, the rows of Cr atoms are positioned in the middle of the unit cell.
    }
    \label{fig:MM_distribution}
\end{figure*}

The fact that the magnetic moment on Ru is induced by the localized moment of Cr can be further proved by analyzing the distribution of the local magnetic moment over the crystal.
In Fig.~\ref{fig:MM_distribution} we plot such a distribution for the two lowest energy structures (but the same observation holds for other structures).
In both cases, Cr atoms form a plane of Cr-O bonded octahedra, and the farther away the Ru atoms are from this plane, the lower the local magnetic moment is, and for the farthest Ru atoms, the local moment effectively collapses to zero.
The same result is observed in Ref.~\onlinecite{wang_emergent_2023}, where the supercells of \ce{RuO2} with 25\% Cr doping were analyzed: as can be seen from Supplementary Table 1 reported therein, for b-axis doubled cell, the local magnetic moment decreases in magnitude the farther the Ru atoms is from Cr.

In conclusion, our calculations show that if doping the parent \ce{RuO2} with Cr would lead to the emergence of the itinerant magnetism, the obtained magnetic ground state would be altermagnetic, when the doping level is above 17\% of Cr.
However, our supercell calculations at 20\% Cr concentration show that Ru atoms remain essentially non-magnetic, while Cr has a sizable magnetic moment, making the itinerant magnetism ansatz not applicable in this case.
Thus, our calculations suggest the need for reinterpretation of the experimental data, which is provided in the following section.

\subsection{Previous experimental results}
In this section we summarized the results reported in Ref.~\onlinecite{wang_emergent_2023}, giving them, if needed, a different angle in view of the latest experimental findings.

\subsubsection{Curie-Weiss susceptibility}
Ref.~\onlinecite{wang_emergent_2023} reports a Curie-Weiss susceptibility that has several
interesting, but not so far addressed features.
First and foremost, one has to recall that nobody has ever observed either a N\'eel transition or a  Curie-Weiss behavior in 
pure \ce{RuO2}: it was speculated in Ref.~\onlinecite{berlijn_itinerant_2017} that the 
N\'eel temperature $T_N$ must be above 400 K; normally (barring magnetically frustrated systems), introducing magnetic ions 
cannot reduce $T_N$.
Yet, in Ref.~\onlinecite{wang_emergent_2023} a magnetic transition was observed at $T\sim 50$ K (for Cr concentration $x=0.2$).
This unequivocally proves, in agreement with Refs.~\onlinecite{smolyanyuk_fragility_2024,hiraishi_nonmagnetic_2024,kessler_absence_2024}, that without Cr, \ce{RuO2} is not magnetically ordered and does not even have localized magnetic moment.

\begin{figure*}
    \includegraphics[width=\textwidth]{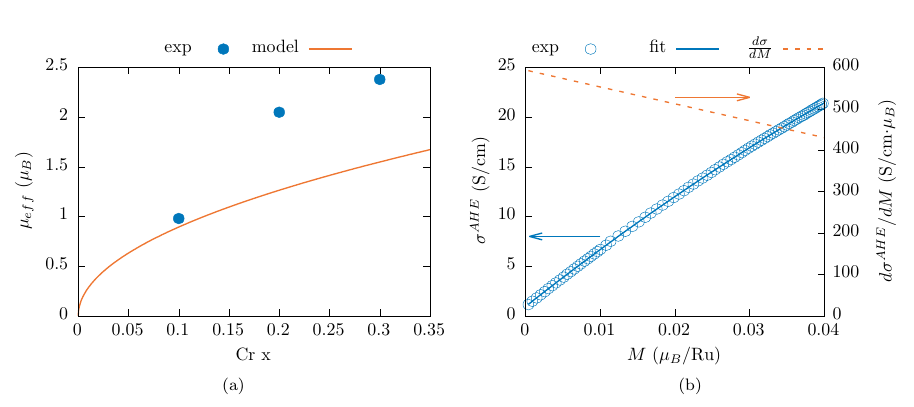}
    \caption{(a) Effective Curie-Weiss moments from Ref.~\onlinecite{wang_emergent_2023} compared to a Cr-only model, neglecting Cr-Cr clusterization, and taking Cr to be a spin 1 ion.
    (b) Anomalous Hall conductivity measured in  Ref.~\onlinecite{wang_emergent_2023} and its derivative (dashed orange line) with respect to the magnetic moment.
    The blue solid line is a quadratic fit to the experimental points.}
    \label{fig:model}
\end{figure*}

Assuming, in the first approximation, that Ru ions even in the doped samples remain nonmagnetic (or, more correctly, as discussed in the previous section, simply screen Cr), effectively renormalizing the effective Cr moment in the Curie-Weiss law, we see that the total susceptibility on the per-Ru basis, plotted in Fig.~2b in Ref.~\onlinecite{wang_emergent_2023}, is proportional to $x$, and the effective moment, on the same basis, to $\sqrt{x}$. As Fig.~\ref{fig:model}a shows, this model correctly reproduces the sublinear trend but underestimates the response at higher concentrations. This is readily understood from the fact that at $x=0.2$ and $x=0.3$ there is a high probability of formation of sizeable \ce{CrO2} clusters.
Indeed, for $x=0.2$ and assuming the random occupation, the probability of an individual Cr to have three or more Cr neighbors is 20\%, for $x=0.3$, 45\%. While it is difficult to say at which size a Cr cluster starts behaving as a single ferromagnetic particle, given that \ce{CrO2} is a strong ferromagnet ($T_C=386$ K), it is liable to happen, and then in the temperature 
interval $T\lesssim 150$~K, used in Ref.~\onlinecite{wang_emergent_2023} some fraction of Cr 
will be bound in such ferromagnetic clusters. Their effect can be estimated from an assumption that $all$ Cr are bound in clusters of the size $n$. Their magnetic moment would be $nM_{Cr}$, and their concentration $x/n$.
Thus, their Curie-Weiss effective moment will be roughly $\mu^2\approx (n M_{Cr})^2 x/n$, roughly $\sqrt{n}$ times larger than the effective moment of the same amount of individual Cr moments, $\mu^2\approx M_{Cr}(M_{Cr}+2) x$.
This explains why the experimentally extracted moments are larger than the simple expectation from individual Cr ions
\footnote{To get the effective magnetic moment per Cr one would use the $\mu_{eff} = \sqrt{8 C/x}$ expression, obtaining $\mu_{eff}=3.10$, $4.58$, and $4.35~\mu_B$ for $x=0.1$, $0.2$, and $0.3$.}.

\subsubsection{High-field magnetization}
While the full saturation has not been reached in Ref.~\onlinecite{wang_emergent_2023}, the moment at 7~T was reported to be 0.03, 0.04, and 0.12 $\mu_B$/f.u. for $x=$0.1, 0.2 and 0.3, respectively. Since we know now that this magnetization is nothing but screened Cr moments, this translates into 0.3, 0.2, and 0.4  $\mu_B$/Cr. To appreciate the message, we recall that the Curie-Weiss temperature determined in Ref.~\onlinecite{wang_emergent_2023} for $x=0.2$ was -75~K, corresponding for $M_{Cr}=2.05~\mu_B$ to the field of 55~T, indicating that at $H=7$~T the response is still strongly suppressed by antiferromagnetic correlations among Cr. It is worth noting that the percolation threshold for the bct lattices is about $x=0.24$~\cite{dalton_dependence_1964}, so $x=0.3$, where some ferromagnetism was detected, is beyond the percolation limit, and $x=0.2$ not much below.

\subsubsection{Low-field susceptibility}
The low field susceptibility~\cite{wang_emergent_2023} for $x=0.2$ shows a smooth, slightly sublinear behavior, with $\chi_{110}$ approximately 50\% smaller than $\chi_{001}$, which was interpreted as the magnetic response of an ordered antiferromagnet with an easy axis [110].
Let us see if this interpretation is possible at all.

\ce{RuO2} is a tetragonal material. Correspondingly, if it were magnetically ordered, it would have possessed a uniaxial anisotropy $E_1=K_1M_z^2$, in the second order in the spin-coupling interaction, and an in-plane anisotropy $E_2=K_2 M_x^2M_y^2$, in the fourth order.
Typically, $E_1$ is on the order of 0.1--1 meV, and $E_2$ of 1-10 $\mu$eV.
If $K_1>0$ and $K_2<0$ the material has an easy-plane anisotropy and the easy in-plane direction is [110] then $\chi_{001}=1/2(2J_{eff}-K_1)$, where $J_{eff}$ is the total antiferromagnetic coupling preventing the canting.
Correspondingly, $\chi_{110}$ would be, initially, zero for the domains oriented along [110] and $1/2(2J_{eff}+K_2)$ for those along 
[1$\bar{1}$0].
Assuming an equal concentration, this means that initially the domain-averaged $\chi_{110}$ will be about half of $\chi_{001}$, and then, after some critical field value, will spin-flop into a state with susceptibility slightly larger than $\chi_{001}$.
Even if the spin-flop field is above 7 T and thus unobservable (highly unlikely, given the very small scale of the in-plane anisotropy), the fact that the ratio of $\chi_{001}/\chi_{110}$ is only 1.5, and not 2 is inconsistent with a scenario of magnetic Ru with the easy axis of 110.

\subsubsection{Anomalous Hall Conductivity}
In Ref.~\onlinecite{wang_emergent_2023}, Hall conductivity in the plane, perpendicular to the [110] direction in the field, parallel to this direction, was measured as a function of the magnetic field.
It was interpreted under the assumption that Cr-doped \ce{RuO2} is a canted altermagnet.
The total Hall conductivity was partitioned into three additive terms: the ordinary Hall conductivity $\sigma_H^O\propto H$, the anomalous conductivity coming from canting of Ru moments, $\sigma_H^M\propto M$, and the residual component  $\sigma_H^A$, which was an unknown function of the field. The latter was presumed to be coming from the magnetic domain population disbalance $\delta(H)$, so that  $\sigma_H^A\propto \delta(H)$.

There are several conceptual issues with this partitioning.
First, the anomalous Hall effect in collinear magnets requires spin-orbit coupling and corresponding induced Berry phases.
The resulting Hall resistivity does not have to be linear in magnetization and quite often cannot be partitioned into two additive terms, one for canting and the other for altermagnetism~\cite{liu_multipolar_2024}.
Furthermore, if the latter is just a small correction to the former, as in Ref.~\onlinecite{wang_emergent_2023}, this partitioning is particularly impossible.

\begin{figure*}
    \includegraphics[width=\textwidth]{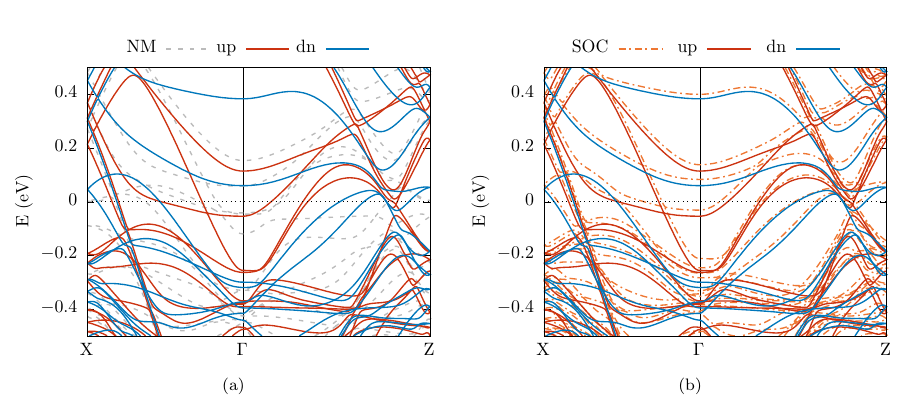}
    \caption{Bandstructure plots for the lowest energy structure among the set of studied supercells:
    (a) comparison between the nonmagnetic (dashed gray line) and spin-polarized FM calculations (solid red and blue lines for projection on spin up and down, respectively) and
    (b) comparison between the FM calculations with (dash-dotted orange line)
    and without SOC contribution.
    Note that the ferromagnetic exchange splitting is larger than SOC.
    }
    \label{fig:bands}
\end{figure*}

Second, the classical linear dependence of $\sigma_H^M$ on $M$ is only correct for ferromagnets in a multidomain state. There is no physical argument for a linear dependence in a canted state, as long as the ferromagnetic exchange splitting is larger than spin-orbit coupling --- which is commonly the case and - as seen from our calculated magnetic band structures without (Fig.~\ref{fig:bands}a) and with spin-orbit coupling (Fig.~\ref{fig:bands}b) - applies also to our Cr-doped \ce{RuO2}.
Moreover, unless the canting sign is correlated with the sense of the altermagnetic order (for whatever reason), the latter does not interact with the field, and the altermagnetic domain $\delta$ does not depend on the field.
If canting does couple with altermagnetism, as it was observed, for instance, in a true altermagnet, MnTe~\cite{Betancourt,Chilcote,kluczyk2023,Nirmal}, then the observed altermagnetic Hall conductivity $\sigma^A$  follows the net magnetization $M$.

It is worth noting that while $\sigma_H^M$ does not have to be linear with $M$, in this experiment it nearly is (cf. Fig.~4b in Ref.~\onlinecite{wang_emergent_2023}).
Fig.~\ref{fig:model}b shows that the measured $\sigma^{AHE}$ is exceedingly well described by a linear function with a small quadratic correction so that there is absolutely no reason to fit the large-$M$ part separately and then subtract the fit from the data.
This is an artificial procedure, and the remainder is an artifact, simply reflecting the small but natural nonlinearity of the regular ferromagnetic AHC~\cite{liu_multipolar_2024}.

To conclude this section, the data presented in Ref.~\onlinecite{wang_emergent_2023} are completely incompatible with a scenario of \ce{RuO2} being a canted altermagnet, and completely compatible with the scenario of random ferromagnetic clusters of \ce{CrO2} embedded in nonmagnetic, but polarizable in a standard way \ce{RuO2}.

\subsection{Conclusions}
In summary, our ab initio calculations suggest that the introduction of Cr dopants into \ce{RuO2} does not lead to the onset of magnetism on Ru atoms, which would have been consistent with a picture of the roughly uniformly hole-doped itinerant magnet.
Indeed, we show that \textit{if} such doping were the case, itinerant (alter)magnetism would have emerged.
Unfortunately, instead, the magnetism in this system is mostly localized and residing on Cr atoms, the magnetization of Ru atoms being just an induced response.
This leads to a different interpretation of the results reported in Ref.~\onlinecite{wang_emergent_2023}.

The fact that the ferromagnetic ground state is preferable in our supercell calculations suggests that it is likely that Cr atoms form ferromagnetic clusters.
They, in turn, are coupled antiferromagnetically.
This would explain the observed in Ref.~\onlinecite{wang_emergent_2023} susceptibility: 
at the high-temperature clusters are of similar size and magnetic moment, thus the susceptibility is Curie-Weiss-like, but at the lower temperature
there is a disbalance, causing the susceptibility to deviate from the Curie-Weiss behavior.
And since clusters themselves are coupled antiferromagnetically, one can observe the antiferromagnetic-like behavior at low temperatures, which might actually be sample-dependent.

\section{Methods}
For the computations, we employed density functional theory (DFT) in the local density approximation (LDA) with the Slater exchange~\cite{dirac_note_1930} and Perdew-Zunger parametrization of Ceperley-Alder Monte-Carlo correlation data~\cite{ceperley_ground_1980, perdew_self-interaction_1981} as implemented in the
\texttt{VASP}~\cite{kresse_ab_1993,kresse_ab_1994,kresse_efficiency_1996,kresse_efficient_1996} package within the projector augmented wave method
(PAW)~\cite{blochl_projector_1994,kresse_ultrasoft_1999};
Ru\_pv, Cr\_pv, and O pseudopotentials were used.
The energy cutoff was set to 700~eV and a regular $\Gamma$-centered mesh with $R_k$ length of 70 was used.

We used the virtual crystal approximation (VCA) as implemented in \texttt{VASP} as well~\cite{eckhardt_indirect-to-direct_2014, bellaiche_virtual_2000}.
The linear interpolation of structures is between experimentally reported structures for \ce{RuO2}~\cite{cotton_crystal_1966, sd_0541479} and \ce{CrO2}~\cite{chamberland_crystal_1967, sd_1102811}.
The reported magnetic moment is a weighted average between the local magnetic moment on Cr and Ru atoms, with the weights being the concentrations of the corresponding atoms.

Symmetry-inequivalent supercells with 5 times the volume of the primitive cell were generated using the \texttt{enumlib} package~\cite{hart_algorithm_2008, hart_generating_2009, hart_generating_2012} producing 52 structures in total and each containing two Cr atoms.
For each cell 4 calculations were performed: one non-magnetic calculation and three spin-polarized ones.
The latter employed three possible spin patterns: ferromagnetic, altermagnetic (note that in this arrangement, two Cr atoms can have the spin aligned in the same direction depending on their atomic position), and where only Cr atoms were initialized with a non-zero magnetic moment, but having opposite directions.

\section{Acknowledgements}
A.S. was supported by the Austrian Science Fund (FWF) through the grant No.~I~6142 (project RECORD).
L.S. acknowledges funding by the Deutsche Forschungsgemeinschaft (DFG, German Research Foundation)-TRR288-422213477 (project A09).
I.M. was supported by the Army Research Office under Cooperative Agreement Number W911NF-22-2-0173. 

The computational results presented have been achieved using the Vienna Scientific Cluster (VSC) and the cluster of the Institute of Solid State Physics of TU Wien.
The crystal structures were plotted using VESTA software~~\cite{VESTA}.
For some of the pre- and post-processing \texttt{alfow}~\cite{AFLOW} and \texttt{vaspkit}~\cite{VASPKIT} tools were employed.
  
This research was funded in whole or in part by the Austrian Science Fund (FWF) [10.55776/I6142].
For open access purposes, the author has applied a CC BY public copyright license to any author accepted manuscript version arising from this submission.

\section{Additional information}
The data needed to reproduce and verify the results presented in this manuscript is publicly available on the TU Wien Research Data repository~\cite{repository}.

\bibliography{bibliography}

\end{document}